\documentclass[rmp,final,a4paper,showkeys,amsmath,twocolumn]{revtex4}
\usepackage{epsfig}
\usepackage{subfigure}

\newcommand{\bm}[1]{\mbox{\boldmath $#1$}}

\begin{document}

\title{On the attractors 
  of two-dimensional Rayleigh oscillators including noise}

\author{Udo Erdmann} 
\email{udo.erdmann@physik.hu-berlin.de}
\author{Werner Ebeling}

\affiliation{Institut f\"ur Physik, Humboldt-Universit\"at zu Berlin,
  Newtonstr. 15, 12489 Berlin, Germany}

\date{\today}

\begin{abstract}
  We study sustained oscillations in two-dimensional oscillator systems driven
  by Rayleigh-type negative friction. In particular we investigate the
  influence of mismatch of the two frequencies. Further we study the influence
  of external noise and nonlinearity of the conservative forces. Our
  consideration is restricted to the case that the driving is rather weak and
  that the forces show only weak deviations from radial symmetry. For this
  case we provide results for the attractors and the bifurcations of the
  system.  We show that for rational relations of the frequencies the system
  develops several rotational excitations with right/left symmetry,
  corresponding to limit cycles in the four-dimensional phase space. The
  corresponding noisy distributions have the form of hoops or tires in the
  four-dimensional space. For irrational frequency relations, as well as for
  increasing strength of driving or noise the periodic excitations are
  replaced by chaotic oscillations.
\end{abstract}



\maketitle


\section{Introduction}
\label{sec:introduction}

The first comprehensive theory of nonlinear oscillations was developed by Lord
Rayleigh in the years 1883--1894 and is represented in his pioneering book
\cite{Ra94}. The present state of art of nonlinear dynamics was deeply
influenced by the pioneering work of Leonid Shilnikov \citep[see][and
references therein]{Sh97,ShShTuCh98a,ShShTuCh98b} which is reflected e.g. in
the books of \citet{An95,AnAsNeVaSchi02}.
In the last decade many investigations were devoted to the stochastic dynamics
of dissipative hamiltonian systems \citep{Kl94,AnAsNeVaSchi02}.

In this paper we extend previous studies on rotational excitations of
two-dimensional oscillators driven by Rayleigh-type negative friction
\cite{ErEbSchiSchw99,ErEbAn00}. The study of driven oscillatory modes on a
plane has interesting applications for modeling animal mobility all the way
from microorganisms like bacteria and {\em Dictyostelium discoedium} slime
mold \citep{CziBeCoVi96,RaNiSaLe99,BeCoLe00} upto flocks of birds
\citep{ToTu95,WeMaClAlJi01}, schools of fish
\citep{PaViGr02,InKa00,Ni94,HuBaSiMa04}, swarms of {\em Daphnia}
\citep{OrBaCaMo03,ErEbSchiOrMo03} or wildebeest \citep{ToBe04}. For example we
introduced in earlier work the general idea of stochastically moving species,
active Brownian particles.  We want to recall this approach which will be used
later on. Active Brownian particles are Brownian particles with the ability to
take up energy from the environment and use it for the acceleration of motion.
Simple models composed of active Brownian particles were studied in many
earlier works \citep[e.g.][]{Kl94,SchiGr93,HeMo95}.

A specific problem we would like to address here is: what is the consequence of
broken radial symmetry (asymmetry of the conservative forces). In particular
we are interested in the problem: Can spontaneous rotations be stopped by
certain amount of asymmetry?  In contrast to previous studies
\cite{ViCzBeCoSh95,CziBeCoVi96,Al96,ShiSuMiHaSa96,CziStaVi97,LeRaCo00,GrChaTu01,HuBaSiMa04,GrCh04}
the self-propelling feature is modeled here by active Brownian particles with
negative friction \cite{EbSchwTi99,SchwEbTi98,ErEbSchiSchw99,SteEbCa94} which
are able to convert stored internal energy into motion. In this paper, we will
study only the motion, especially rotational ones, in external fields on a
two-dimensional plane.

The article is organized as follows. In Section~\ref{sec:dynam-equat-two} we
introduce the equations of motion, the pumping by negative friction and
outline the basic dynamics of our model including Langevin equations. In
Section~\ref{sec:attr-line-oscill} we outline our previous studies of
rotationally symmetric external potentials and discuss the attractors for
rational frequency relations. In Section~\ref{sec:bifurc-analys-line} we give
an analysis of frequency mismatch and of Arnold-type bifurcations of linear
oscillators as a function of the mismatch.  Further we study in
Section~\ref{sec:nonl-oscill-with} the limit cycle attractors of nonlinear
oscillators with radial symmetry. In Section~\ref{sec:influence-noise} we
investigate the influence of noise on the attractors and show that noise leads
to a broadening on the line-attractors.

\section{Dynamic equations for two-dimensional oscillators}
\label{sec:dynam-equat-two}

The dynamics of the systems studied here is based on Langevin equations, known
from the theory of conventional Brownian motion \citep{La08,HaTaBo90}. For the
two-dimensional space we get four first order coupled differential equations
in the phase space $\{x_1,x_2,v_1,v_2\}$:
\begin{subequations}
  \label{2dmotion}
  \begin{eqnarray}
  \dot{x}_1 &=& v_1 \\
  \dot{x}_2 &=& v_2 \\
  \dot{v}_1 &=& \frac{1}{m}\frac{\partial U}{\partial x_1} 
  -\gamma(v_1,v_2) v_1 + \sqrt{2D}\xi_1(t) \\
  \dot{v}_2 &=& \frac{1}{m}\frac{\partial U}{\partial x_2} 
  -\gamma(v_1,v_2) v_2 + \sqrt{2D}\xi_2(t)\,.
\end{eqnarray}
\end{subequations}
In this dynamics we assumed three kinds of forces in the dynamics:
\begin{enumerate}
\item Conservative external forces generated by the potentials $U(x_1,x_2)$,
\item nonlinear dissipative forces modeled by the friction $(-\bm{v}
  \gamma(\bm{v}))$,
\item stochastic forces assigned by $\xi(t)$.
\end{enumerate}
Several linear and nonlinear conservative forces will be introduced and
studied in the following sections. The dissipative forces are modeled by the
friction function $\gamma(\bm{v})$. This function is the source of
dissipative interactions with the surrounding. In equilibrium the friction is
passive
\begin{equation}
  \gamma(\bm{v}) = \gamma_0 = {\rm const.}
\end{equation}
We will consider here in more detail active friction modeled by the classical
Rayleigh law \citep{Ra94}:
\begin{equation}
  \gamma(\bm{v}) = \gamma(v_1,v_2) = \alpha - \beta(v_1^2  + v_2^2)\,.
  \label{Rayleigh}
\end{equation}

The stochastic forces are modeled by white Gaussian noise $\xi$ with
vanishing mean and 
\begin{equation} 
  \langle\xi(t)\xi(t')\rangle = \delta(t-t')
  \label{noise}
\end{equation} 
and scaled with strength $D$. In equilibrium and in case of passive friction
following \citet{Ei56} one gets an energy balance between the strength of the
stochastic force, $D$, and the passive friction acting on the object. It is
expressed by the simple fluctuation-dissipation relation
\begin{equation}
  D = \frac{\gamma_0 \theta}{m}
  \label{theta}
\end{equation}
where $\theta = k_B T$ is a measure for the temperature.

\section{Attractors for linear oscillators driven by Rayleigh friction}
\label{sec:attr-line-oscill}

We will study in this section two-dimensional linear oscillators described by
the potential
\begin{equation}
\label{eq:pot}
U(x_1,x_2) = \frac{m}{2}(\omega_1^2 x_1^2 + \omega_2^2 x_2^2)
\end{equation}
which are driven by Rayleigh-type negative friction as in
Eq.~(\ref{Rayleigh}). It is well known since \citet{Ra94} that in the
one-dimensional case the system possesses a limit cycle corresponding to
sustained oscillations with the energy $E_0=\alpha/\beta$. The two-dimensional
case is much more complicated. \citet{ErEbSchiSchw99} have shown for the
symmetrical case with $\omega_2=\omega_1$, that two limit cycles in the
four-dimensional phase space are developed. The projection of these sustained
oscillations on the $\{v_1,v_2\}$-plane and on the $\{x_1,x_2\}$-plane are
circles
\begin{subequations}
  \begin{eqnarray}
    v_1^2 + v_2^2 &= v_0^2 &= \frac \alpha \beta\\
    x_1^2 + x_2^2 &= r_0^2 &= \frac{v_0}{\omega_1}
  \end{eqnarray}
\end{subequations} 
The limit cycle energy is
\begin{equation}
  E_0 = \frac{mv_0^2}{2} +\frac{m\omega_1^2}{2}\, r_0^2 \;.
\end{equation}

\citet{EbSchwTi99} have shown, that any initial value of the energy converges
(at least in the limit of strong pumping) to
\begin{equation}
  H\longrightarrow E_0 = m v_0^2
\end{equation}
This corresponds to an equal distribution between kinetic and potential energy
i.e. both parts contribute the same amount to the total energy. The motion on
the limit cycle in the four-dimensional space may be represented by the four
equations
\begin{subequations}
  \label{4d-explizit}
  \begin{eqnarray}
    x_1 &=& r_0 \cos(\omega_0 t + \phi_0) \\
    v_1 &=& - r_0 \omega_0 \, \sin(\omega_0 t + \phi_0)\\ 
    x_2 &=& r_0 \sin(\omega_0 t + \phi_0) \\
    v_2 &=& r_0 \omega_0 \, \cos(\omega_0 t + \phi_0)
  \end{eqnarray}
\end{subequations}
The angular frequency follows by estimations of the time the particle needs
for one period moving on the circle of radius $r_0$ with constant speed $v_0$:
\begin{equation}
  \omega_0 = \frac{v_0}{r_0} = \omega_1=\omega_2 \;.
\end{equation}
This means, the particle rotates even at strong pumping with the frequency
given by the linear oscillator frequency $\omega_1=\omega_2$.  The trajectory
defined by Eqs.~(\ref{4d-explizit}) is an exact solution of the dynamic
equations describing the first (forward) limit cycle. The shape of the
trajectory is like a hoop in the four-dimensional space. Most projections to
the two-dimensional subspaces are circles or ellipses however there are two
subspaces namely $\{x_1,v_2\}$ and $\{x_2,v_1\}$ where the projection is like
a rod \cite{ErEbSchiSchw99}.  Reversing the initial velocities of the system,
a second limit cycle can be obtained. This limit cycle forms also a hula hoop
which is different from the first one. However both limit cycles have the same
projections on the $\{x_1,x_2\}$ and on the $\{v_1,v_2\}$-plane. The
projection to the $\{x_1,x_2\}$-plane has the opposite sense of rotation in
comparison with the first limit cycle. The projections of the two hoops on the
$\{x_1, x_2\}$-plane or on the $\{v_1, v_2\}$-plane are two-dimensional rings.
The hoops intersect perpendicular $\{x_1, v_2\}$- and $\{x_2, v_1\}$-planes.
The projections to these planes are rod-like and the intersection manifold
with these planes consists of two ellipses located in the diagonals of the
planes \citep{ErEbSchiSchw99}.

So far we repeated known results for the symmetrical case. Dynamical systems
with radial symmetry are degenerate and structurally unstable in the
mathematical context. From the physical viewpoint, radial symmetry is a
special situation, i.e. the gravitational field of point masses or the Coulomb
field for charges has strict radial symmetry. Therefore radial symmetry holds
also for a two-dimensional mass-point pendulum. In real physical systems the
radial symmetry is in general broken, e.g. a real pendulum in the earth field
has no strict radial symmetry. Thus the oscillator with radial symmetry can
only be considered as a particular case of corresponding real system which
often has some asymmetry. Thus our motivation is, to investigate the
consequences of broken radial symmetry and deviations from linearity on the
generation of oscillatory modes.

\begin{figure}[htbp]
  \begin{center}
    \epsfig{file=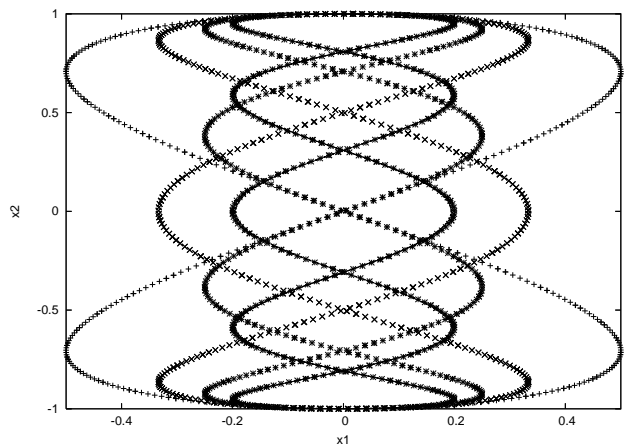, width=8.5cm}
    \caption{\label{AS3} Projections of asymmetric limit cycles to the 
      $\{x_1,x_2\}$-plane for several cases of rational relations between the
      two frequencies $m:n = 2,3,4,5$ (analytical approximation for the
      attractors (Eqs.~(\ref{periodsol}) and (\ref{analattr})).}
  \end{center}
\end{figure}

\begin{figure}[htbp]
  \begin{center}
    \epsfig{file=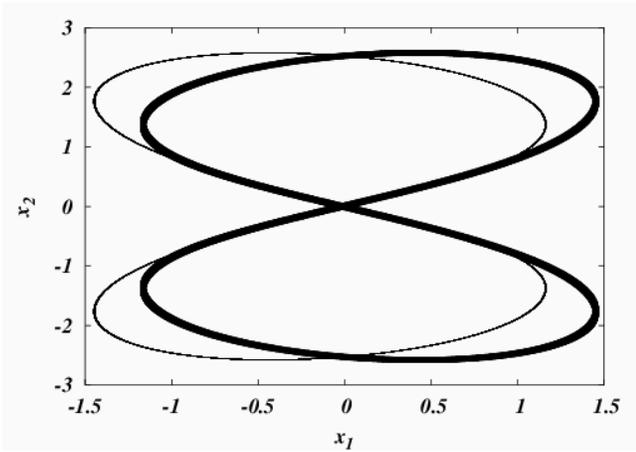, angle=-90, width=8.5cm}
    \caption{\label{d2xx} Projections of the two limit cycles to the 
      $\{x_1,x_2\}$-plane corresponding to $m:n=2$ resonance obtained from
      simulations (Rayleigh law: $\alpha=5$, $\beta=1$).}
  \end{center}
\end{figure}

\begin{figure}[htbp]
  \begin{center}
    \epsfig{file=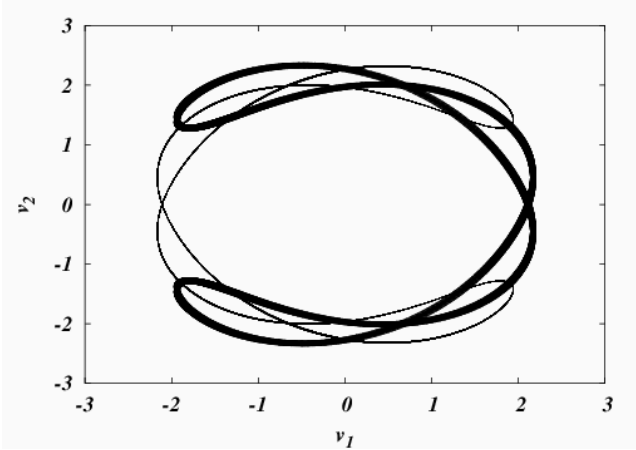, angle=-90, width=8.5cm}
    \caption{\label{d2vv} Projections of the two limit cycles to the 
      $\{v_1,v_2\}$-plane corresponding to $m:n=2$ to the $\{v_1,v_2\}$-plane
      obtained from simulations (same parameters as in Fig.~\ref{d2xx}).}
  \end{center}
\end{figure}

\begin{figure}[htbp]
  \begin{center}
    \epsfig{file=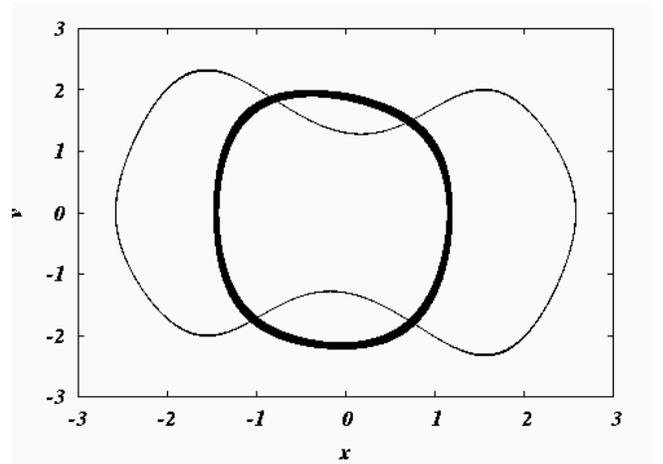, angle=-90, width=8.5cm}
    \caption{\label{d2xv} Projections of the two limit cycles to the 
      $\{x_2,v_2\}$-plane for $m:n=2$ and respectively obtained from
      simulations (same parameters as in Fig.~\ref{d2xx}). Clockwise and
      counterclockwise limit cycle have practically the same projections.}
  \end{center}
\end{figure}

First we are going to study systems with frequency mismatch $\omega_2 \ne
\omega_1$. In the conservative case $\gamma =0$ and $D=0$ we find a dense set
of exact solution for the deterministic dynamics given by
\begin{subequations}
  \label{periodsol}
  \begin{eqnarray}
    x_1(t) &=& A_1 \cos(\omega_1 t + \Phi_1)\\
    v_1(t) &=& - A_1 \omega_1 \sin(\omega_1 t + \Phi_1),\\
    x_2(t) &=& A_2  \sin(\omega_2 t + \Phi_2)\\
    v_2(t) &=& A_2 \omega_2 \cos(\omega_2 t + \Phi_2)\,,
  \end{eqnarray}
\end{subequations}
where $A_i$ and $\Phi_i$ are given by the (arbitrary) initial conditions.  Let
us study now the driven case with negative friction according to the Rayleigh
law (\ref{Rayleigh}) without noise. Then for $\alpha > 0$ the above periodic
solution (\ref{periodsol}) would remain to be a solution, if
\begin{equation}
  v_1^2 + v_2^2 = v_0^2 = \frac{\alpha}{\beta}
\label{condition}
\end{equation}
is fulfilled. In the case of symmetrical oscillators $\omega_1=\omega_2 =
\omega_0$ we could fulfill this condition in an exact way by a special choice
of the amplitudes:
\begin{equation}
  A_1 = A_2 = r_0 = \frac{v_0}{\omega_0}
\end{equation}
This means, we find instead of a dense set of exact solutions just one
attracting solution (which still is exact).
This periodic and stable exact solution represents a limit cycle corresponding
to a circular path even at strong pumping with the frequency given by the
simple harmonic oscillator frequency $\omega_0$. However in the case
$\omega_1\ne \omega_2$ the situation is much more difficult than in the
symmetrical case. First we are looking for approximate stable solutions in the
case of rational relations of the frequencies $\omega_2=n\omega_0$ and
$\omega_1 = m\omega_0$ with $n, m = 1,2,\ldots$. The condition
(\ref{condition}) is in average over one period fulfilled, if
\begin{equation}
  A_1^2 \omega_1^2 = A_2^2 \omega_2^2 = v_0^2
\end{equation}
This way we get the stable amplitudes and phases
\begin{subequations}
  \begin{eqnarray}
    A_1 = \frac{v_0}{\omega_1}\,; \qquad A_2 = \frac{v_0}{ \omega_2}\\
    \Phi_1= -\frac \pi 2 \,; \qquad \Phi_2=+\frac \pi 2
\end{eqnarray}
\label{analattr}
\end{subequations}
We will show that these approximative solutions describe at least for small
relations $n:m$ again a pair of forward/backward limit cycles. By introducing
the approximation \ref{analattr} into Eq.~(\ref{periodsol}) we find for the
$n>m$ the following analytical approximation for the limit cycles:
\begin{subequations}
  \label{lcnm.eq}
  \begin{eqnarray}
    x_1(t) &=& r_0 \sin(m\omega_0 t)\\
    v_1(t) &=& r_0 m\omega_0 \cos(m\omega_0 t)\\
    x_2(t) &=& r_0  \cos(n\omega_0 t)\\
    v_2(t) &=& -r_0 n\omega_0 \sin(n\omega_0 t)\,.
  \end{eqnarray}
\end{subequations}
The curves obtained by this approximation for $n:m = 2,3,4,5$ are shown in
Fig.~\ref{AS3}.
In the analytical approximation given above, the forward limit cycles and the
backward limit cycles have identical projections on the $\{x_1,x_2\}$-plane.
However our analytical formula (Eqs.~(\ref{periodsol}) and (\ref{analattr}))
is only a rough approximation, as we will demonstrate by simulations. We show
in Figs.~\ref{d2xx} and \ref{d2vv} the result of simulations for the two
attractors in the case $m:n=2$.
Clearly, the two attractors have different projections on the
$\{x_1,x_2\}$-plane (see Fig.~\ref{d2xx}) as well as on the
$\{v_1,v_2\}$-plane (see Fig.~\ref{d2vv}). As we see, the projection differ in
amplitude and phase a little bit from the analytical approximation; in
particular clockwise and counterclockwise limit cycles are shifted in the
projections. It is interesting to note that the projections to the
$\{x_1,v_1\}$-plane and to the $\{x_2,v_2\}$-plane shown in Fig.~\ref{d2xv}
are equal for the clockwise and the counterclockwise limit cycles.
  
\begin{figure}[htbp]
  \begin{center}
    \epsfig{file=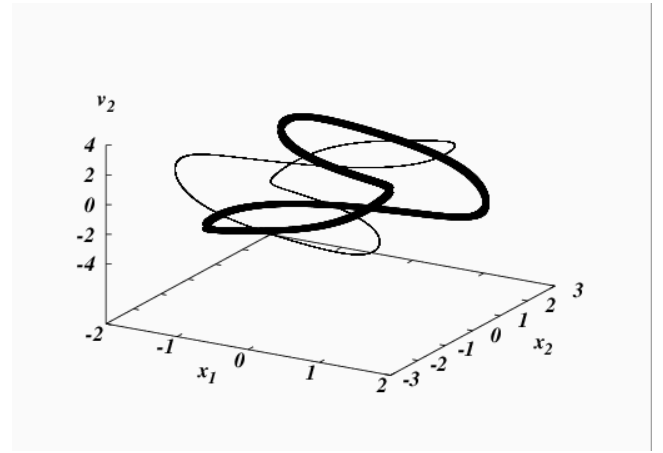, angle=-90, width=8.5cm}
    \caption{\label{d2xxv} Projections of the two limit cycles to the 
      $\{x_1,x_2,v_2\}$-plane corresponding to the $m:n=2$-resonance obtained
      from simulations (same parameters as in Fig.~\ref{d2xx}).}
  \end{center}
\end{figure}

The rather complex winding structure of the attractor for $n=2$ in the
four-dimensional phase space can be guessed by looking at projections on
three-dimensional subspaces as demonstrated in Fig.~\ref{d2xxv}.

\begin{figure}[htbp]
  \begin{center}
    \epsfig{file=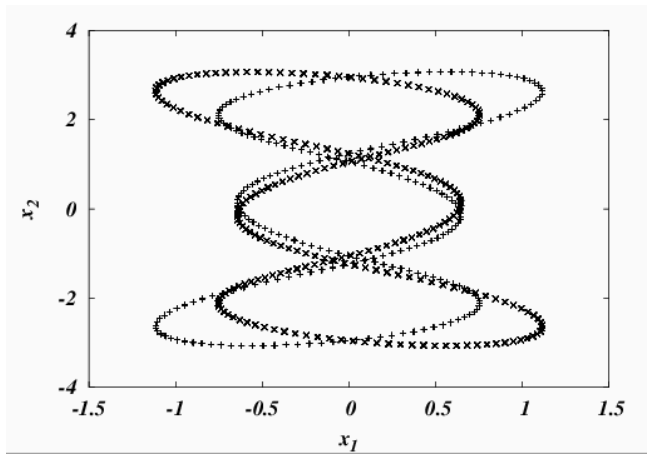, angle=-90, width=8.5cm}
    \caption{\label{d3xx} Projections of the two limit cycles to the 
      $\{x_1,x_2\}$-plane corresponding to the $m:n=3$-resonance obtained from
      simulations ($\omega_1=2.7$, $\omega_2=1$, all other parameters as in
      Fig.~\ref{d2xx}).}
  \end{center}
\end{figure}

\begin{figure}[htbp]
  \begin{center}
    \epsfig{file=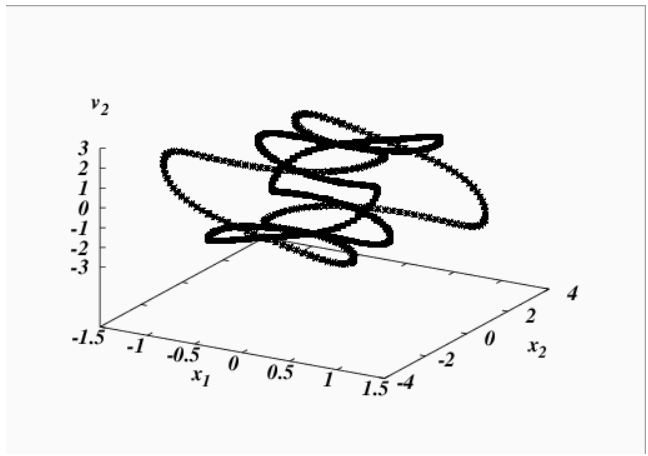, angle=-90, width=8.5cm}
    \caption{\label{d3xxv} Projections of the two limit cycles for $m:n=3$ to
      the $\{x_1,x_2,v_2\}$-plane obtained from simulations.}
  \end{center}
\end{figure}

The results obtained from simulations for $n=3$ are shown in Figs.~\ref{d3xx}
and \ref{d3xxv}. We see that the attractors obtained from simulations are more
complex than the from the nice and rather symmetrical curves obtained from the
analytical approximation for the case $n=3$ which were presented in
Fig.~\ref{AS3}.  With increasing rational $n$ the attractor fills more or less
dense a rectangular region. For irrational values of $\omega_2/\omega_1$ the
trajectories are always dense in a nearly rectangular region of the coordinate
space. This case will be studied below.

\section{Bifurcation analysis of the linear oscillators with frequency
  mismatch}\label{sec:bifurc-analys-line}

Our study was limited so far to small rational frequency relations $m:n$ and
small or moderate strength of driving (small positive values of the
bifurcation parameter $\alpha$). In order to introduce the consequences of a
an irrational frequency mismatch between the partial subsystems in a more
general setting we write the expression for the potential Eq.~(\ref{eq:pot})
as follows:
\begin{subequations}
  \label{ellipse}
  \begin{eqnarray}
    U(x_1,x_2)&=&\frac{m}{2}\, \omega_0^2(x_1^2+ \Delta^2 x_2^2), \\
    \Delta&=&\frac{\omega_1}{\omega_2}; \qquad \omega_0=\omega_1.
  \end{eqnarray}
\end{subequations}
In the general case the potential $U(x_1,x_2)$ has an elliptic, slightly
extended shape, the asymmetry is measured by the parameter $\Delta$. For
irrational values of $\Delta$ the solutions in the conservative case
$\alpha=\beta=0$ fill densely a box in the phase space. In the general driven
case the deterministic part of Eq.~(\ref{2dmotion}) may be written as
\begin{subequations}
  \label{eq:lang-detune}
  \begin{eqnarray}
    \dot{x}_1&=&v_1 \\
    \dot{v}_1&=&\left[\alpha-\beta\left(v_1^2
        +v_2^2\right)\right]v_1-\omega_0^2 x_1\\
    \dot{x}_2&=&v_2 \\
    \dot{v}_2&=&\left[\alpha-\beta\left(v_1^2
        +v_2^2\right)\right]v_2-\omega_0^2 \Delta^2 x_2
  \end{eqnarray}
\end{subequations}
The system (\ref{eq:lang-detune}) is structurally stable or is one of the
common propositions according to V.~I.~Arnold's nomenclature
\citep{Ar65}. It can be interpreted as a model for one oscillator on the
plane or as a model for two interacting linear oscillators.

\begin{figure}[htbp]
  \centering
  \epsfig{file=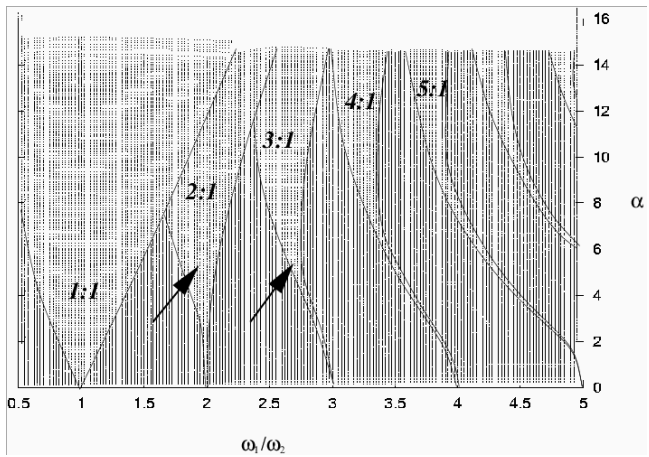,angle=-90, width=8.5cm}
  \caption{Stability regions (Arnold tongues) of the $m:n$ resonances. Within
    the light gray areas the limit cycles as described above are stable. The
    arrows show the parameter set of Figs.~\ref{d2xx}-\ref{d3xxv}.}
  \label{fig:arnold}
\end{figure}

To understand the dynamics of system (\ref{eq:lang-detune}) in the general
case, we introduce a complex variable $z=x_1+jx_2$. Let us first turn again to
the symmetric case where $\omega_1=\omega_2=\omega_0$. Then from
(\ref{eq:lang-detune}) follows
\begin{equation}
  \label{eq:complex}
  \ddot{z}-\beta\left(\frac{\alpha}{\beta}-|\dot{z}|^2\right)\dot{z}
  +\omega_0^2z=0.
\end{equation}
Equation (\ref{eq:complex}) has periodic solutions of the form:
\begin{equation}
  \label{eq:sol_complex}
  z(t)=z \exp(\pm j\omega_0t)=|z| \exp(j\Phi)\,\exp(\pm j\omega_0 t),
\end{equation}
where the phase $\Phi$ takes any value in the interval $[0,2\pi]$. When we
consider the symmetric case, we have an infinite number of periodic solutions
(see Eq. (\ref{eq:sol_complex})). However, linear analysis cannot yield
information about their stability. In numeric calculations, we can detect
several limit cycles each of them possessing its own type of symmetry. As by
\citet{ErEbAn00} has been investigated, the stability of periodic solutions
can can be predicted by the calculation of the Floquet multipliers of the fixed
points within the Poincar{\`e} section of the limit cycle. When there is a
detuning ($\Delta \neq 1$), only two limit cycles, representing clockwise and
counterclockwise rotations remain stable as has already been found by
\citet{ErEbSchiSchw99}. All other solutions have at least one multiplier
leaving the unit circle on the complex plane. The two limit cycles
representing the rotations of a particle in the coordinate space are situated
within the first Arnold tongue whereas outside this region the existence of
the cycles vanishes and the particles move irregularly. Increasing the
detuning $\Delta$ any further new stability regions appear which represent the
period-$n$ cycles of the oscillators as have been shown in
Figs.~\ref{AS3}-\ref{d3xxv}. The parameter region where stability of periodic
motion can be observed is shown in Fig.~\ref{fig:arnold}.

\section{Nonlinear oscillators with radial symmetry}
\label{sec:nonl-oscill-with}

In the present section we will discuss several extensions of the theory
developed in the previous section to nonlinear oscillators. However, for
simplicity, we will restrict our study to systems with rotational symmetry,
$U=U(r)$ with $U(0)=0$ and $U(r)-$ monotonically increasing. For the general
case of radially symmetric but anharmonic potentials $U(r)$ the equal
distribution between potential and kinetic energy is violated. In other words
the relation $mv_0^2 = m\omega_0^2 r_0^2$ which leads to $\omega_0 = v_0/r_0 =
\omega$ is no more valid. This relation has to be replaced by the more general
condition that on the limit cycle the attracting radial forces are in
equilibrium with the centrifugal forces. This leads to
\begin{equation}
\frac{m v_0^2}{r_0} = |U'(r_0)| 
\end{equation} 
If $v_0$ is given, the equilibrium radius may be found from the implicit
relation
\begin{equation}
\label{impl}
v_0^2 = \frac{r_0}{m} |U'(r_0)|  
\end{equation} 
Then the frequency of the limit cycle oscillations is given by
\begin{equation}
\omega_0^2 = \frac{v_0^2}{r_0^2} = \frac{|U'(r_0)|}{m r_0} 
\end{equation} 
In the case of linear oscillators this leads us back to previous result given
in Sec.~\ref{sec:attr-line-oscill}. For the case of quartic oscillators
\begin{equation}
U(r) = \frac{k}{4} r^4  
\end{equation} 
we get the limit cycle frequency
\begin{equation}
\omega_0 = \frac{k^{1/4}}{v_0^{1/2}} 
\end{equation} 
The explicite solution (\ref{4d-explizit}) remains valid, i.e. we find again
an exact analytical description of the pair of limit cycles. For monotonically
increasing potentials there exist just one stable radius $r_0$. If the
equation (\ref{impl}) has several solutions, the dynamics might be much more
complicated. An interesting application of the theoretical results given above
is the case of Coulomb forces \cite{SchiEbEr05}. Another possible application
is the following: Let us imagine a system of Brownian particles which are
pairwise bound by a Lennard-Jones-like potential $U(|r_1 - r_2|)$ to
dumb-bell-like configurations.  Then the motion consists of two independent
parts: The free motion of the center of mass, and the relative motion under
the influence of the potential. As a consequence, the center of mass of the
dumb-bell will make a driven Brownian motion but in addition the dumb-bells
are driven to rotate around there center of mass. What we observe then is a
system of pumped Brownian molecules which show driven translations with
respect to their center of mass. On the other side the internal degrees of
freedom are also excited and we observe driven rotations. \citet{ErEbAn00}
have shown that this approach can be extended to systems of many particles in
the sense that as far as the mean field of the interaction potential can be
approximated like potentials of the shape of Eq.~(\ref{eq:pot}), rotations are
going to be stable within the first Arnold tongue (Fig.~\ref{fig:arnold}).  In
this way we have shown that the mechanisms described here may be used also to
excite the internal degrees of freedom of Brownian molecules.

\section{The influence of noise}\label{sec:influence-noise}

The main effect of noise is the spreading of the deterministic attractors.
Let us consider here only the case of radially symmetric potentials $U(r)$
which have a mini\-mum $U=0$ at $r=0$ and are monotonically increasing with
$r$. Then as shown above, the system has two limit cycles in the
four-dimensional space which are hoop-like and have projections at the
$\{x_1,x_2\}$-space which are circles with the radius $r_0$. This radius is
determined by the equilibrium between centripetal and centrifugal forces for
right/left rotations on the circle with the radius $r_0$.  Including
stochastic effects the two hoop-like limit cycles are converted into
distributions looking like two embracing hoops with finite size, which for
strong noise convert into two embracing tires in the four-dimensional phase
space.

\begin{figure*}[htbp]
  \begin{center}
    \subfigure[$\Delta=1$]{\label{distrib_a}
      \epsfig{file=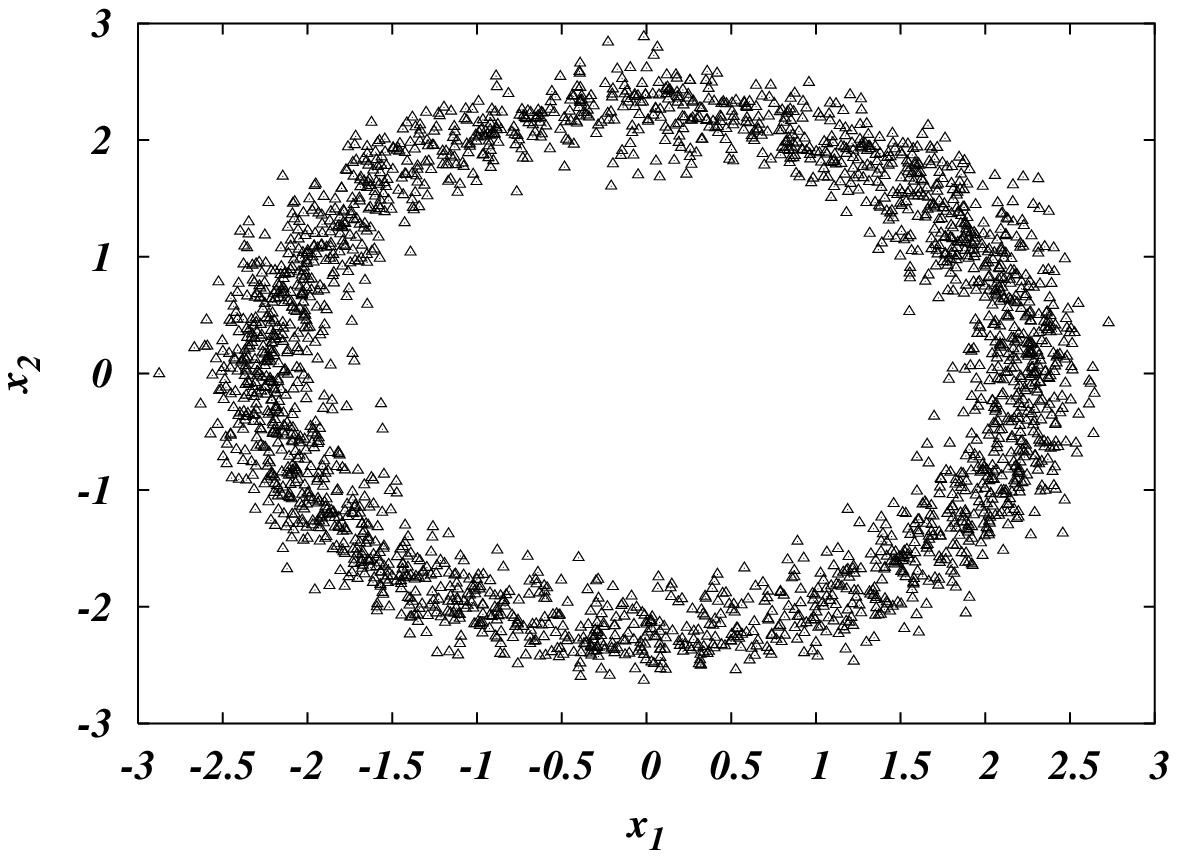, width=8cm}}
    \subfigure[$\Delta=1$]{\label{distrib_b}
      \epsfig{file=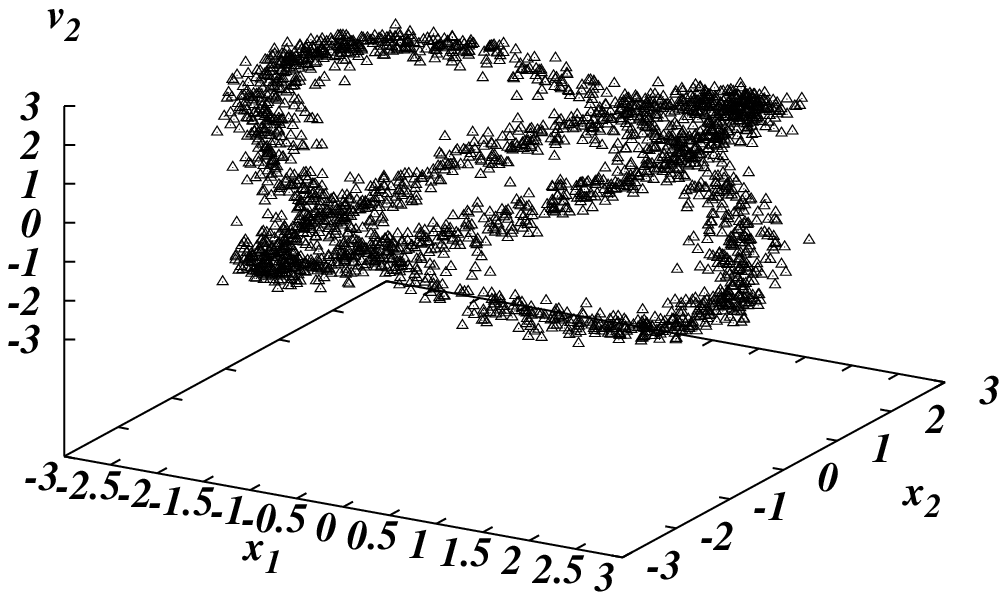, width=8cm}}
    \subfigure[$\Delta=2$]{\label{distrib_c}
      \epsfig{file=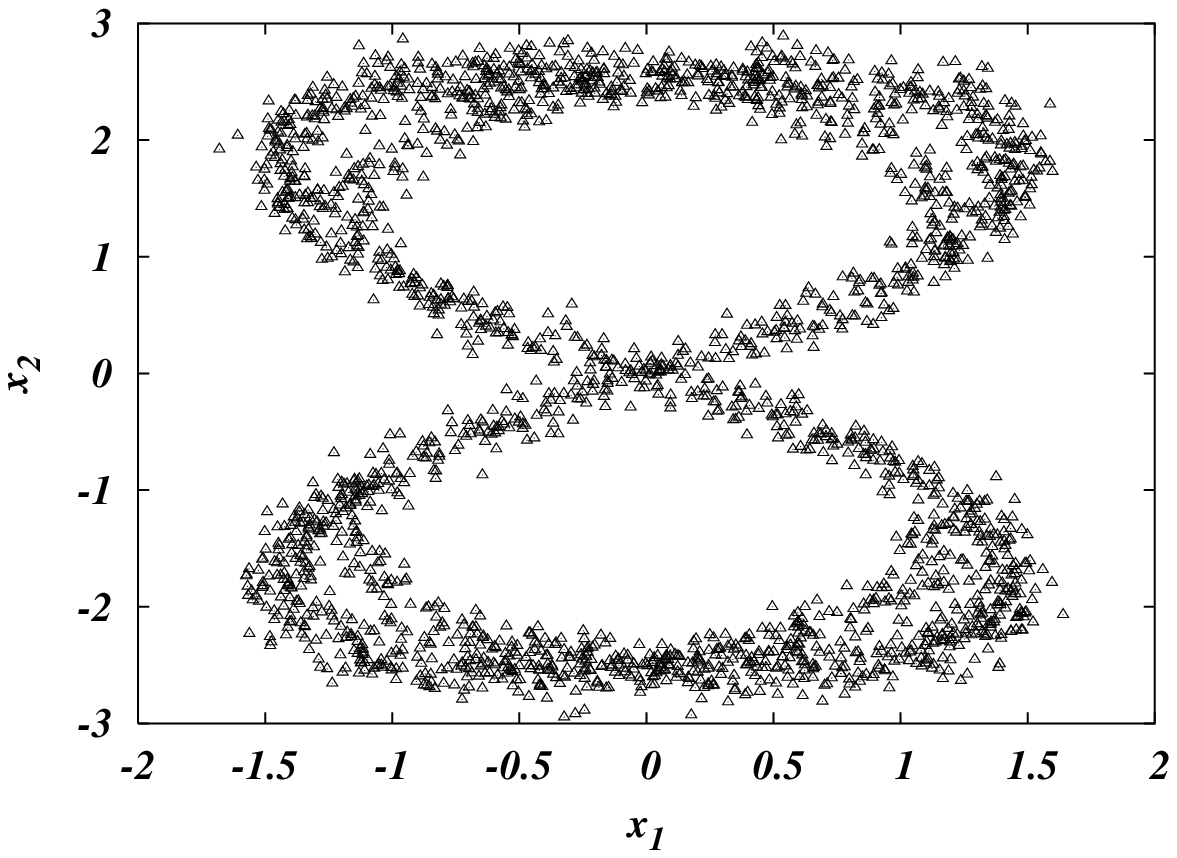, width=8cm}}
    \subfigure[$\Delta=2$]{\label{distrib_d}
      \epsfig{file=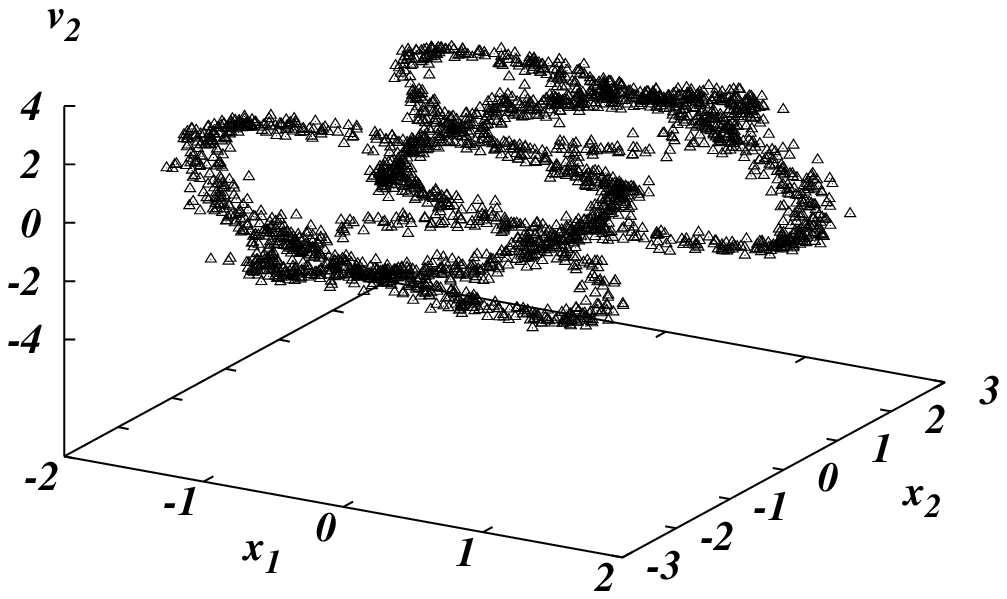, width=8cm}}
    \subfigure[$\Delta=3$]{\label{distrib_e}
      \epsfig{file=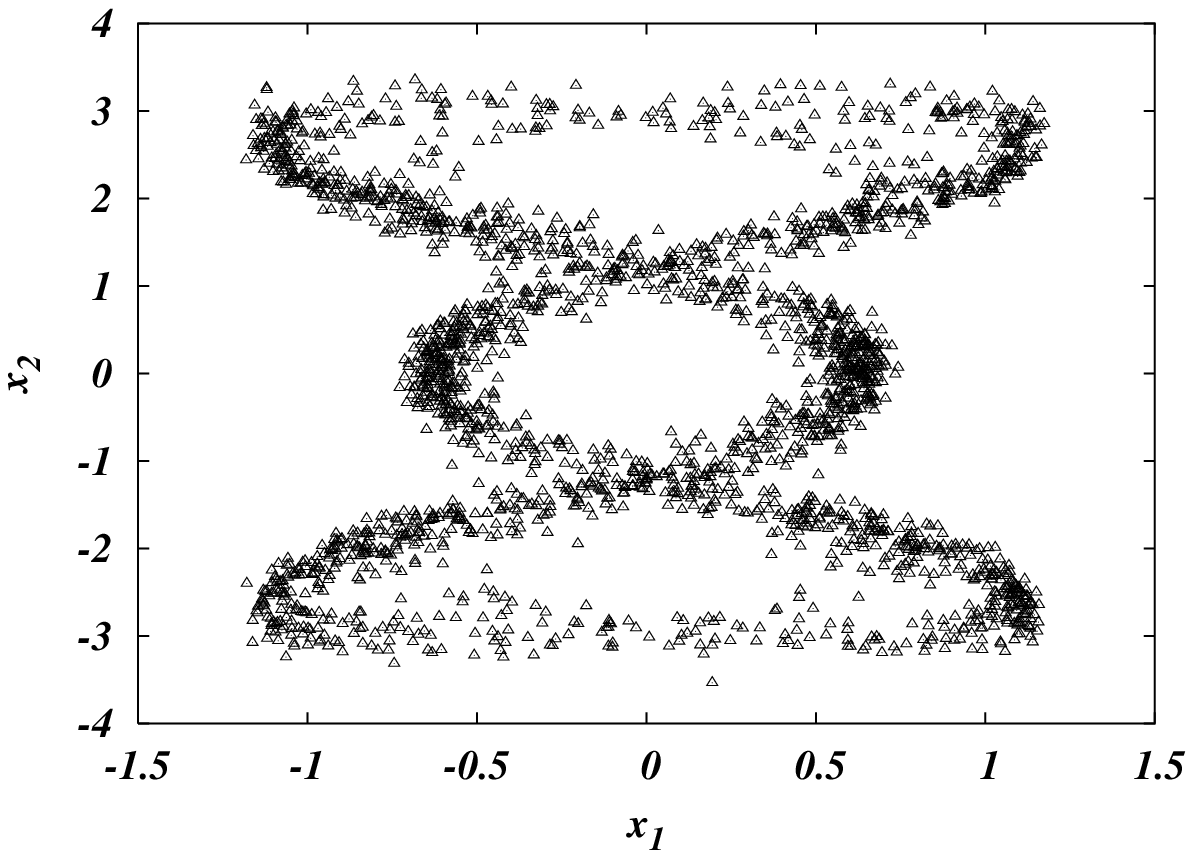, width=8cm}}
    \subfigure[$\Delta=3$]{\label{distrib_f}
      \epsfig{file=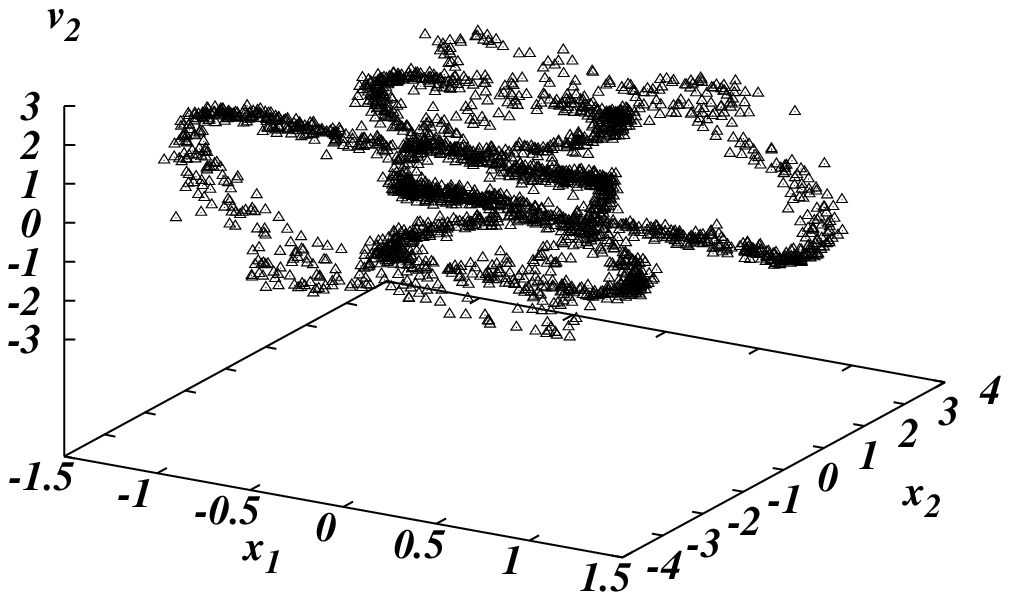, width=8cm}}
    \caption{\label{distrib} Distributions according to simulations of 2500
      Active Brownian particles on limit cycles in symmetric (a-b) and
      asymmetric (c-f) parabolic potentials. We show projections of
      simulations on the $\{x_1,x_2\}$-plane and on the subspace
      $\{x_1,x_2,v_2\}$. Parameters: $\alpha = 5, \beta=1, D=0.01$.}
  \end{center}
\end{figure*}

The transformation of the limit cycles into tires can easily be obtained from
simulations including white additive noise. Several results for both symmetric
($\Delta=1$) and asymmetric ($\Delta\neq 1$) parabolic potentials are
demonstrated in Fig.~\ref{distrib}. 
The probability distributions may be obtained as solutions of the
Fokker-Planck equation for the probability distribution $P(\bm{r},\bm{v},t)$
\cite{Kl94}
\begin{eqnarray}
  \label{fpe-or}
  \frac{\partial P}{\partial t}  &=& 
  - \bm{v}\, \frac{\partial P}{\partial \bm{r}} - \nabla U(\bm{r}) \,
  \frac{\partial P}{\partial\bm{v}}\\ &&+\frac{\partial}{\partial \bm{v}}
  \left\{\gamma(\bm{v})\,\bm{v}\, P + D\, 
    \frac{\partial P}{\partial \bm{v}}\right\} \nonumber
\end{eqnarray}
To find explicite solutions of Eq.~(\ref{fpe-or}) is a very difficult task
\cite{Kl94,AnAsNeVaSchi02}. Only in the force-free case $U=0$, the solution is
elementary. The stationary solution reads for the Rayleigh-model
\cite{ErEbSchiSchw99,ErEbAn00}:
\begin{equation}
  \label{RayHelm_stat}
  P_0(\bm{v}) = C \exp\left[\frac{\alpha\bm{v}^2}{2 D} 
  \left(1-\frac{\bm{v}^2}{2 v_0^2}\right)\right]\,.
\end{equation}
The shape of this velocity distribution Eq.~(\ref{RayHelm_stat}) can be seen
in \citep[e.g.][]{ErEbAn00}. The bifurcation to a limit cycle at the
transition from negative to positive $\alpha$ occurs for the noisy system as a
qualitative change of the shape of the distribution from a Maxwell-like to a
hat-like shape. It is obvious that the system above the bifurcation point is
far from equilibrium and shows a permanent active motion of the particles.
Applying similar arguments to the stochastic motion in confining potentials we
expect that the two hoops are converted into a distribution with the
appearance of two embracing hoops with finite size, which for strong noise
converts into two embracing hoops in the four-dimensional phase space
\citep[see][for details]{ErEbSchiSchw99}. In order to obtain the explicit form
of the distribution, we may introduce the amplitude-phase representation
\begin{subequations}
  \begin{eqnarray}
    x_1 &=& \rho \cos(\omega_0 t + \phi) \\
    v_1 &=& -\rho \omega_0 \sin(\omega_0 t + \phi)\\
    x_2 &=& \rho \sin(\omega_0 t + \phi) \\
    v_2 &=& \rho \omega_0 \cos(\omega_0 t + \phi)\,,
\end{eqnarray}
\end{subequations}
where radius $\rho$ and phase $\phi$ are slow and fast stochastic variables
respectively. By using the standard procedure of averaging with respect to the
fast phases we obtain for the Rayleigh-model of pumping the following
distribution of the radii:
\begin{equation}
P_0(\rho) \simeq  \exp{\left[\frac {\alpha v_0^2}{D} 
    \left(\frac{\rho}{r_0}\right)^2
    \left(1 -\frac{\rho^{2}}{2 r_0^2} \right)\right]}\;.
\end{equation}

\begin{figure}[htbp]
  \centering
  \begin{minipage}{8cm}
    \epsfig{file=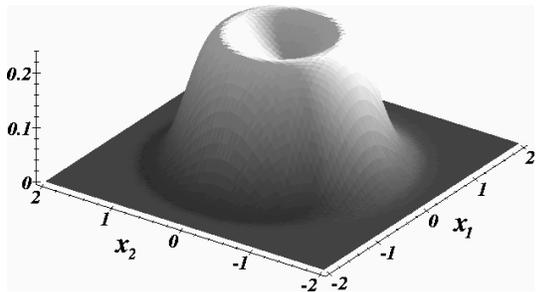,angle=-90, width=7cm}
  \end{minipage}
  \caption{\label{P_rho} Probability distribution $P_0(\rho)$ for a 
    radially symmetric potential in the coordinate space $\{x_1,x_2\}$. The
    parameters $\alpha, r_0, v_0,D$ are set equal to 1.}
\end{figure}

This distribution is (in the present approximation) universal and valid for
any radially symmetric potential of the type specified above.  We see that the
probability crater is determined by the two deterministic limit cycles (see
Fig.~\ref{P_rho}). The velocity distribution given by Eq.~(\ref{RayHelm_stat})
remains to be exact for all radially symmetric potentials. The full stationary
probability in the four-dimensional space has the form of two hula hoop
distributions \cite{SchiEbEr05,DeZh04}. The projections of the distribution
onto the $\{x_1,x_2\}$-plane and to the $\{v_1,v_2\}$-plane are
two-dimensional rings (see also Fig.~\ref{distrib}(a-b)). As in the
deterministic case the hula hoop distribution intersects perpendicularly the
$\{x_1,v_2\}$-plane and the $\{x_2,v_1\}$-plane. Again, the projections to
these planes are rod-like, and the intersection manifold with these planes
consists of two ellipses located in the diagonals of the planes
\cite{ErEbSchiSchw99,SchiEbEr05}. In the deterministic case one of two the
rotational motions within the confining potential is excited, this rotation
remains a stable solution of the trajectory of one particle. To this rotation
belongs a certain value of the angular momentum. For non-vanishing stochastic
perturbations, the particle is able to cross the separatrix between the two
rotational modes (limit cycles). Due to this ability of the particles one can
observe, sometimes, an inversion of the angular momentum of the particle
\cite{ErEbAn00}. Note that one initial condition will be sufficient to rotate
either clockwise or counterclockwise (see Fig.~\ref{distrib_b}).

\section{Conclusions}
\label{sec:conc}

We study here two-dimensional oscillations with nonlinear Rayleigh-type
active friction, extending earlier work to
\begin{enumerate}
\item oscillations with frequency mismatch
\item oscillations with radially symmetric nonlinear attractive forces.
\end{enumerate}
We investigate several right/left symmetric pairs of limit cycles
corresponding to 1:1, 2:1 and 3:1 resonances and their location in the
stability regions (Arnold tongues). We show that in the presence of noise the
limit cycles are converted into hoops in the four-dimensional space and give
analytical estimates for the probability distributions of the coordinates and
velocities. Thinking of applications especially for coherent behavior in
animal motion on could think of looking for interacting particles where the
interaction could be approximated (in mean field approximation) as a deviated
parabolic potential. In extention to \citep{ErEbAn00} on should investigate
more the question if there would be any coherent motion which could be located
within the higher Arnold tongues. How would the motion of a swarm of particles
look than?

\begin{acknowledgments}
  The authors acknowledge the fruitful discussions with V.S. Anishchenko
  (Saratov University) and L. Schimansky-Geier (Humboldt-University Berlin,
  Germany). Furthermore the support by J. Dunkel (Humboldt-University Berlin,
  Germany) and A. Neiman (Ohio University) with some numerics is acknowledged.
  The work is partly supported by the Collaborative Research Center ``Complex
  Nonlinear Processes'' of the German Science Foundation (DFG-Sfb 555).
  
  It is a deep pleasure for the authors to express their sincere gratitude to
  Leonid P. Shilnikov for many suggestions and advice. As a Humboldt-fellow,
  L.P.S. spent during the last 3-4 years many months in Berlin sponsored by
  the Humboldt-Foundation. Many seminars and discussions with L.P.S. in small
  circles at Humboldt-University Berlin were extremely useful and inspiring to
  us.  His very clear way of pointing out the characteristic features of
  nonlinear dynamics in combination with wonderful friendly personal relations
  influenced deeply our work. Thanks, Leonid and all good wishes to your
  jubilee.
\end{acknowledgments}
\bibliographystyle{apsrmp}
\bibliography{bib/allgemein,bib/brown,bib/bakterien,bib/cells,bib/coherent,bib/erdmann,bib/gbt,bib/plasma,bib/schleimpilze}

\end{document}